# On-chip interference of single photons from an embedded quantum dot and an external laser

N. Prtljaga[1,a], C. Bentham[1,b], J. O'Hara[1,b], B. Royall[1], E. Clarke[2], L. R. Wilson[1], M. S. Skolnick[1], and A. M. Fox[1]

[1]*Department of Physics and Astronomy, University of Sheffield, Sheffield S3 7RH, United Kingdom*
[2]*Department of Electronic and Electrical Engineering, University of Sheffield, Sheffield S1 3JD, United Kingdom*

In this work, we demonstrate the on-chip two-photon interference between single photons emitted by a single self-assembled InGaAs quantum dot and an external laser. The quantum dot is embedded within one arm of an air-clad directional coupler which acts as a beam-splitter for incoming light. Photons originating from an attenuated external laser are coupled to the second arm of the beam-splitter and then combined with the quantum dot photons, giving rise to two-photon quantum interference between dissimilar sources. We verify the occurrence of on-chip Hong-Ou-Mandel interference by cross-correlating the optical signal from the separate output ports of the directional coupler. This experimental approach allows us to use classical light source (laser) to assess in a single step the overall device performance in the quantum regime and probe quantum dot photon indistinguishability on application realistic time scales.

Two-photon interference between photons originating from different quantum emitters is at the heart of proposals for linear optics quantum computing[1,2]. Furthermore, interference between single photons and coherent states is of considerable interest for a number of practical implementations in quantum communications and quantum key distribution[3–5]. The essential ingredient for quantum interference to take place is that the interfering photons are mutually indistinguishable in all observable degrees of freedom[1]. The degree of indistinguishability between the incoming photons is commonly quantified by performing a Hong-Ou-Mandel (HOM) experiment[6]. For many practical applications, it is highly desirable that the generation of single photons and the manipulation of the photon states all take place on a single chip[7–9], opening routes to scalable quantum photonics. An open question is to what extent photon indistinguishability is maintained when the quantum emitter is embedded within realistic photonic circuits. In this Letter, we demonstrate on-chip two-photon interference of dissimilar sources using a single self-assembled InGaAs quantum dot (QD) monolithically integrated with a beam-splitter and an attenuated external laser. The observation of two-photon interference demonstrates that useful levels of photon indistinguishability are maintained within a photonic environment, confirming the suitability of the GaAs material platform for quantum information processing.

---

[a] Author to whom correspondence should be addressed. Electronic mail: n.prtljaga@sheffield.ac.uk

[b] C. Bentham and J. O'Hara contributed equally to this work.

The device consists of a thin layer (140 nm) of GaAs with QDs embedded in the centre which is formed by etching the waveguides in this layer and by removing the sacrificial layer beneath, leaving it suspended. Measurements are performed in an exchange gas cryostat at 4.2 K using a confocal microscope system with four independent optical paths (two excitation and two collection). The photoluminescence (PL) signal is generated using a Ti:Sapphire continuous wave (CW) laser emitting at 840 nm for wetting layer excitation. Transmission and laser/QD interference measurements are performed using a single mode tunable laser (30 kHz linewidth). For detection, we use single-photon avalanche photodiodes or a charge coupled device camera. The PL is spectrally filtered with two spectrometers set to a spectral bandwidth of 90 µeV, and in the case of high resolution spectra and laser-dot detuning adjustments, a scanning Fabry-Perot interferometer (0.3 µeV resolution) is used.

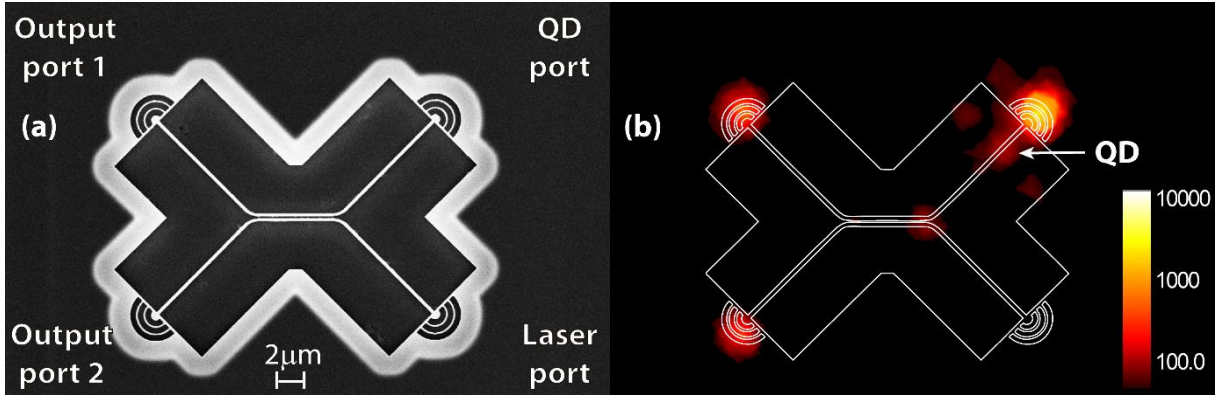

FIG. 1. (Colour online) (a) Typical scanning electron microscope (SEM) image of a four port directional coupler (seen as bright lines in the image) used in this work. The device has been underetched and is suspended. (b) Logarithmic colour scale PL map of the tested device with an overlaid device contour. Spectrally filtered single QD PL is used as an internal light source illuminating the device. The approximate QD position within the directional coupler arm, from which PL is detected, is also indicated.

An example of the directional coupler device used to first combine and then split photons is shown in Fig. 1(a). The main advantages of this type of on-chip beam-splitter are the simple design, ease of fabrication, low-losses and well understood behaviour at the single photon level[10–12]. The device consists of two single-mode waveguides that are brought into close proximity, allowing for evanescent light coupling[13]. The waveguide dimensions (140 nm high and 280 nm wide) were chosen in order to ensure operation in a single polarisation (TE)[10]. Optical simulations are performed using a commercial-grade eigenmode solver[14]. Coupled mode theory is then used to determine the optimum coupling length ($L=7$µm) and separation (70 nm) between the two waveguides for 50:50 operation[13].

We characterise the wavelength dependence of the splitting ratio (S.R.) of the device by performing transmission measurements with an external tunable laser Fig. 2(a). The measured spectral dependence agrees well with the theoretical dependence obtained from coupled mode theory for the target device design ($L=7$ µm, waveguide separation 70 nm)[10]. At the emission wavelength of the QD, the device performance can be approximated as an $R$(reflection):$T$(transmission)= 55:45 beamsplitter.

For the laser-dot interference experiment, we select a spectrally isolated QD (see Fig. 2(b)). This emission line shows a linear power dependence and lacks any resolvable fine structure (<2 µeV), suggesting it is due to charged exciton recombination. The linewidth of this transition has been determined by high-resolution spectroscopy to be 11 µeV, corresponding to a coherence time, $\tau_c$, of 120 ps. While most of the QD lines show larger linewidths, on average at least one QD of similar linewidth can be found per device. The best linewidth measured in devices from this sample is 8 µeV under wetting layer excitation.

The QD PL signal is collected from the output couplers and filtered independently in each collection path, both spectrally and spatially. The orientation of the grating output couplers at the end of each waveguide differs by 90°, providing us with mutually orthogonal linearly polarised signals. Thus, photons coming from opposite output ports are also fully distinguishable in polarisation once they leave the on-chip beam-splitter.

The QD embedded within the input arm of the directional coupler couples efficiently to the propagating mode of the single mode waveguide[10]. This is clearly visible in a PL map of the device (see Fig 1(b)) obtained by raster scanning the collection across the device whilst spectrally filtering at the QD wavelength.

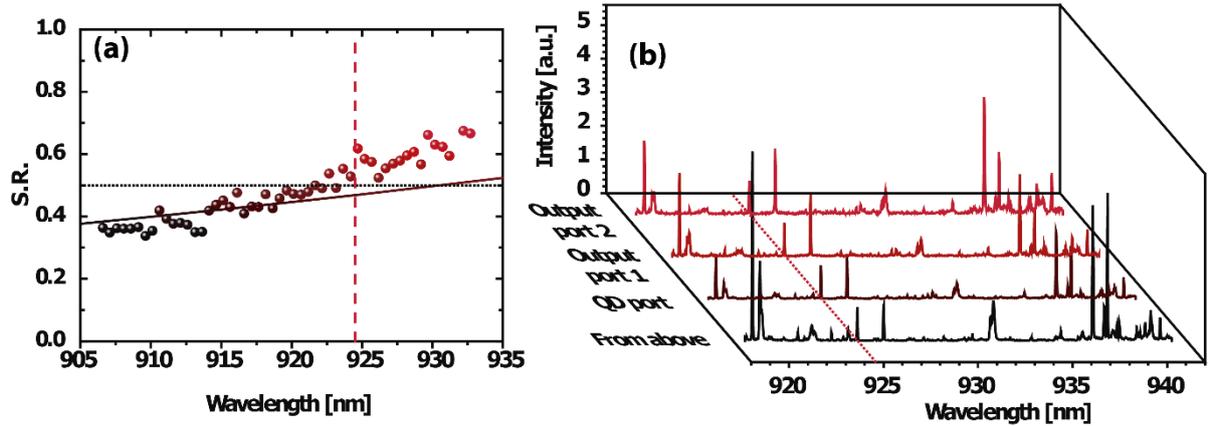

FIG. 2. (Colour online) (a) Measured wavelength dependence (symbols) of the on-chip beam-splitter splitting ratio (S.R.). S.R. increases towards longer wavelength (symbol colour fades from black to red). Small oscillations are due to residual noise and weak Fabry-Perot oscillations from the waveguide ends. The solid line is the theoretical wavelength dependence obtained from coupled mode theory for the nominal device design. The vertical red dashed line indicates the QD emission wavelength. The horizontal black dotted line indicates a splitting ratio of 0.5. (b) PL spectra (solid lines of different colours) of the QD under wetting layer excitation as seen from above the waveguide and various device ports. The red dotted line indicates the emission line used in this work. All spectra are normalised to the intensity of the investigated emission line.

We verify the single-photon nature of the emission from this QD by performing an on-chip Hanbury-Brown Twiss (HBT) experiment, which consists of cross-correlating the photons from the output ports when only the QD signal is passing through the device. The corresponding normalised trace without background subtraction is reported in Fig. 3(a). By deconvolving the experimental data with the temporal response of our detection system ($R_f(\tau)$, Gaussian full width half maximum FWHM=874±4 ps) we obtain $g_{HBT}^{(2)}(0)$=0.06±0.01 ($g_{RAW}^{(2)}(0)$= 0.175±0.010 for the raw data). We also verify that $g_{HBT-Laser}^{(2)}(t)$ for the external laser remains Poissonian at all times ($g_{HBT-Laser}^{(2)}(t)$=1) when passed through the device (not shown).

The two-photon interference takes place when an attenuated laser signal, tuned to resonance with the QD emission line, is added to the other input arm of the directional coupler. The two-photon interference visibility between dissimilar sources, where one source is anti-bunched and the other is Poissonian, is ultimately limited by the Poissonian nature of the second source. According to the theoretical model developed by Legero et al.[15], used in this work, this limit is solely dependent on QD/laser intensity ratio ($\eta/\alpha^2$) for the case of infinitely fast detectors[15,16]. For the more realistic case of a finite temporal response of the detection system ($R_f(\tau)\neq 0$) observable visibility will also depend on the $\tau_c$ and $R_f(\tau)$. For the measurements presented here (see Fig. 3(b)) we maintain the QD/laser intensity ratio ($\eta/\alpha^2$) of 1. This limits the maximum obtainable visibility of two-photon interference (*visibility_HOM*) to ~ 67% for the case of infinitely fast detectors[15,16]. For larger ratios the measurement time increases significantly as it scales with a square of the count reduction for the CW excitation used. This would have a significant impact on the duration of the experiments as we already drive the QD at least an order of magnitude below the saturation value to avoid the power induced line broadening. As discussed below, for smaller ratios the total visibility is decreased.

Since the measurement systems can drift/misalign within the time window of the measurement (days), traces for both, non-interfering ("off") and interfering ("on") cases are acquired by detuning the probe laser off and on the dot frequency every 30 minutes. As the device supports only a single polarisation, the probe laser photons are made distinguishable from the dot photons by frequency detuning the probe laser from the dot by 29 µeV. Both the QD excitation and the external probe laser are actively frequency and power stabilised for the duration of the measurements. As the external laser properties are known and carefully controlled and the $R_f(\tau)$ is fixed, this experimental configuration probes the QD photon indistinguishability over the time scale of measurement (days)[17,18].

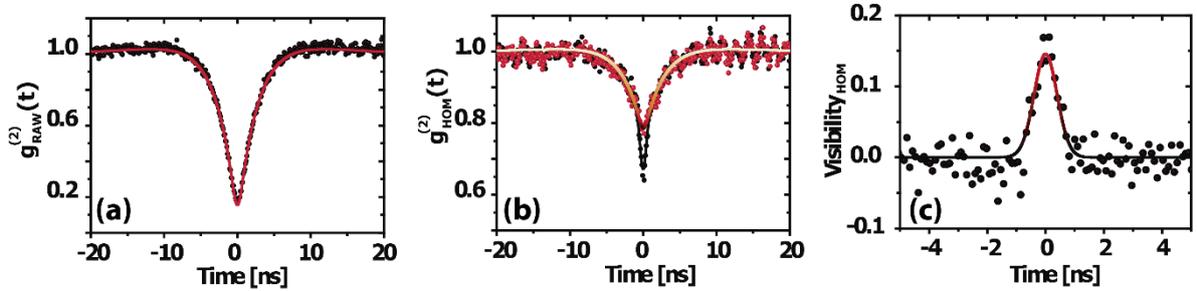

**FIG. 3. (Colour online) (a) Normalised second order correlation function for the QD measured through the on-chip beam-splitter (symbols). The red continuous line is a fit to data taking into account the time response of the measurement system. (b) Cross-correlated signal from the device output ports when the laser is tuned on (black symbols) and off the QD frequency (red symbols). Small oscillations in the experimental data are due to electronic noise in the photon counting system. Corresponding continuous lines are theory predictions. (c) Measured two-photon interference visibility (symbols) and theory (continuous line).**

Results of these measurements are reported in Fig. 3(b). When QD and laser frequencies coincide, i.e. their photons are made indistinguishable ("on" case), the dip in the correlation trace is more pronounced with respect to the "off" case within a time window whose width is determined by the QD coherence time, $\tau_c$. This is a clear signature of two-photon quantum interference. The usual figure of merit used to characterise the magnitude of this effect is the two-photon interference visibility (*visibility_HOM*). *visibility_HOM* is defined here in analogy with the definition from Ref. 15:

$$visibility_{HOM}(\tau) = \frac{g^{(2)}_{OFF}(\tau) - g^{(2)}_{ON}(\tau)}{g^{(2)}_{OFF}(\tau)}, \qquad (1)$$

$$g^{(2)}_{HOM}(\tau) = \left(\frac{\left(\frac{\eta}{\alpha^2}\right)^2 g^{(2)}_{HBT}(\tau) + \left(\frac{\eta}{\alpha^2}\right)\left(\frac{T^2+R^2}{TR} - 2\cos(\Delta E \cdot \tau/\hbar)e^{-\frac{|\tau|}{\tau_c}}\right) + 1}{\left(\frac{\eta}{\alpha^2}\right)^2 + \left(\frac{\eta}{\alpha^2}\right)\frac{T^2+R^2}{TR} + 1}\right) \otimes R_f(\tau), \quad (2)$$

where $g^{(2)}_{OFF}(\tau) = g^{(2)}_{HOM}(\tau, \Delta E = 29\ \mu eV)$ and $g^{(2)}_{ON}(\tau) = g^{(2)}_{HOM}(\tau, \Delta E = 0\ \mu eV)$, and $\Delta E$ is the frequency detuning. The non-zero background and the imperfect beamsplitter performance have also been accounted for here, although their influence is almost negligible for the present case. We obtain a raw visibility of 15 % (Fig. 3(c)), limited by the detector time response, $R_f(\tau)$ and $\tau_c$. The simulated behaviour which includes the system temporal response [15,16,19–21], and which uses parameters obtained from independent linewidth and HBT measurements (Fig. 3(a)), agrees very well with the measured data (Fig. 3(b) and 3(c)). At lower QD/laser intensity ratios of $(\eta/\alpha^2)$ =0.5 and $(\eta/\alpha^2)$ =0.25 (not shown), we measure visibilities of 11% and 8% respectively, again in good agreement with our theoretical model. This implies that the absolute spectral position, single photon purity, coherence time and consequently indistinguishability of photons[17] from this device embedded QD is maintained over the course of days and well described by the model[15].

In conclusion, we have fabricated an on-chip 50:50 beam-splitter with a monolithically integrated QD. We use this device to combine the photons from an external attenuated laser with the photons originating from a QD embedded within the device. By performing correlation measurements on the device when both laser and QD signals are present, we demonstrate on-chip two-photon interference between two dissimilar sources. This experimental approach allows us to use a conventional light source (laser) to assess in a single step the overall device performance in the quantum regime and the stability of QD photon indistinguishability on application realistic time scales. One possible application we envisage for the experimental approach described here is the wafer scale testing of future integrated quantum optical logic gates. This work furthermore paves the way towards demonstration of linear quantum optical circuits with integrated deterministic quantum emitters for quantum computation and/or quantum communication.

The authors would like to thank Paul N. Kemp-Russell and the staff of the Physics Workshop at the University of Sheffield for technical help. This work was funded by EPSRC grant number: EP/J007544/1.